# Data-Driven Revenue Management for Air Cargo

Ezgi C. Eren[1] · Jiabing Li

PROS Inc, Suite 600, 3200 Kirby Dr, Houston, TX 77098, USA

It is well-recognized that Air Cargo revenue management is quite different from its passenger airline counterpart. Inherent demand volatility due to short booking horizon and lumpy shipments, multi-dimensionality and uncertainty of capacity as well as the flexibility in routing are a few of the challenges to be handled for Air Cargo revenue management.

In this paper, we present a data-driven revenue management approach which is well-designed to handle the challenges associated with Air Cargo industry. We present findings from simulations tailored to Air Cargo setting and compare different scenarios for handling of weight and volume bid prices. Our results show that running our algorithm independently to generate weight and volume bid prices and summing the weight and volume bid prices into price optimization works the best by outperforming other strategies with more than 3% revenue gap.

*(Data-driven revenue management, air cargo revenue management, heuristic bid price generation, neural network)*

## 1. Introduction

It is widely recognized that Air Cargo faces some unique challenges in revenue management, some of which may be summarized as follows (Rizzo, Lucas, & Kaoudi, 2019; Boonekamp, 2013; Hoffmann, 2013; Dongling, 2010; Mardan, 2010; Popescu, 2006; Huang & Hsu, 2005):

- *Volatile demand:* Short booking horizon combined with lumpy shipments with variable batch size result in inherently volatile demand that is hard to forecast.
- *Multi-dimensional capacity:* Cargo capacity is multi-dimensional considering both weight and volume.

[1] Corresponding author: eeren@pros.com

- *Uncertain capacity:* Unlike passenger airline, the capacity is highly variable and not easily tractable, due to the uncertainty in space occupied by passenger cargo, allotments (long-term contracts) and such. The booked cargo capacity is also uncertain due to the changes in shipment sizes until departure (over-/under-tendering).
- *Flexible routing:* Unlike passenger airline, cargo can be shipped along any route as long as it makes it to the destination withing a certain time window. That introduces another complexity into the problem with multiple routing options to be considered.

Conventional revenue management (RM) is not an ideal fit for Air Cargo, as it relies on predictable demand patterns with stable seasonality and smooth booking curves; and requires long and stable historical data to forecast demand. For Air Cargo which faces demand volatility on an ongoing basis with its so-called "late and lumpy" demand (Mardan, 2010), generating a demand forecast with an acceptable accuracy is often deemed infeasible.

In this paper, we propose a new approach to solve the RM problem that fits industries facing similar challenges as Air Cargo. We apply the methodology introduced in Eren et al. (2024) to the Air Cargo setting. The data-driven approach directly generates marginal opportunity costs (bid prices) using historical data *without any demand forecasting,* which makes it an especially good fit for volatile settings. The robustness of the methodology is comprehensively studied in Eren et al. (2024). In this paper, we focus on aspects related to Air Cargo setting, and specifically the handling of weight and volume dimensions.

The paper is organized as follows: In Section 2, we review the related literature for Air Cargo revenue management and show how our approach fits in. In Section 3, we describe the problem and solution methodology tailored to air cargo setting in detail. In Section 4, we present the simulation studies and results; and we conclude with a summary of contributions and a discussion of potential future work in Section 5.

## 2. Literature Review

We refer the reader to Eren et al. (2024) for an extensive review of data-driven RM, under which the proposed methodology falls. Here, we will review the RM literature on Air Cargo.

There is relatively limited literature on air cargo RM, mostly highlighting the unique challenges of air cargo and the inability to apply passenger RM techniques directly. Huang & Hsu (2005) focus on the impact of supply (capacity) uncertainty on revenue increase gained by RM, modeling a single-leg dynamic program with a single capacity dimension. Their work introduces the complexity of batch arrivals along with supply uncertainty and show the significance of accurate forecast and control of capacity on revenue gain. Popescu (2006) addresses the demand lumpiness by splitting the modeling of the problem by small and large cargo size and are able to better manage the computational efficiency and tractability by doing so. The small cargo problem is modeled by a probabilistic nonlinear program (PNLP) and the large cargo RM problem is solved via dynamic programming. They generate static bid prices from both models and test the frequency of updating bid prices. Dongling (2010) addresses the multi-dimensionality by incorporating both weight and volume into a discrete-time Markov chain model and generating a bid-price based control policy considering both dimensions, albeit statically. Hoffman (2013) models a similar problem and decomposes it to two one-dimensional sub-problems, by using a proration technique to attribute revenue to weight and volume. Another challenging aspect of air cargo problem is the loss of structural properties, such as monotonicity of bid prices in capacity and time dimensions, due to batch arrivals and the constraint of fully accepting or rejecting a shipment. Hoffman (2013) simplifies the problem by relaxing this assumption and allowing for partial acceptance. Boonekamp (2013) also considers a two-dimensional single-leg problem. However, their solution methodology differs from earlier approaches, as they develop a simulation and ex-post optimum based approach to generate bid prices.

The air cargo RM literature mostly converges on the generation of bid-price based or EMSR type policies. However, to the best of our knowledge, there is no study which attempts to generate dynamic bid price policies, where the output is a vector or a function of remaining capacity and can be used to update bid price dynamically as capacity levels change with each booking request. Although the update frequencies are studied (Popescu, 2006), those require re-solution of the model each time an update is needed. As capacity levels fluctuate quite frequently due to short booking horizon and lumpy shipments, a dynamic bid price vector would address the real-time updates as needed. Another aspect of aforementioned studies is that they assume knowledge of demand distributions and/or parameters and mostly focus on optimization methodologies. Studying of uncertainty is mostly confined to the supply aspect as well as over-booking approaches. Given demand forecasting is the main pain point for Air Cargo, the success of optimization methodologies would heavily depend on forecast accuracy. In this paper, we address both aspects by focusing on a data-driven approach which is primarily designed to address demand volatility and generate a dynamic bid price output.

As a novel data-driven approach to generate bid prices, our work is quite different than the methodologies that are traditionally used for RM including air cargo literature as outlined above. Given the increasing need for more advanced RM with the surging demand during, post-pandemic and beyond (Ozden & Celik, 2021), approaches tailored to the challenges of Air Cargo are needed more than ever.

## 3. Problem Description and Methodology

We adapt the problem description and methodology introduced in Eren et al. (2024) to Air Cargo setting, by extending it to handle batch arrivals with a continuous quantity. Before going into the details of the methodology, we begin with the problem description tailored to Air Cargo use case.

### 3.1 Problem Description

Similar to Eren et al. (2024), we will consider a single cargo flight leg across multiple departures over time. We first begin with consideration of a one-dimensional continuous quantity and capacity before we introduce the complexity of handling two-dimensions. $\boldsymbol{P} = \{P^1, P^2, \ldots, P^{|I|}\}$ and $\boldsymbol{T} = \{T^1, T^2, \ldots, T^{|I|}\}$ represent the sequence of vectors corresponding to the booked per unit prices and the time-to-departure for each flight $i \in I = \{1, 2, \ldots, |I|\}$. In addition to booked price and time data, to incorporate batch arrivals with a continuous quantity, we have $\boldsymbol{Q} = \{Q^1, Q^2, \ldots, Q^{|I|}\}$ representing the sequence of vectors that hold the quantity booked (e.g., weight of the shipped cargo). The components of the vectors $P^i = [p_1^i, \ldots, p_n^i, \ldots, p_{N^i}^i]^\top$, $T^i = [t_1^i, \ldots, t_n^i, \ldots, t_{N^i}^i]^\top$, $Q^i = [q_1^i, \ldots, q_n^i, \ldots, q_{N^i}^i]^\top$ hold the price, time, and quantity information for all bookings: For the $i^{th}$ flight, $N^i$ represents the total number of bookings, and $p_n^i$ is the price, $t_n^i$ is the time-to-departure, and $q_n^i$ is the weight shipped for the $n^{th}$ booking on that flight. Note that $t_1^i > t_2^i > \cdots > t_{N^i}^i$ by definition. It is common to record the time-to-departure in an aggregated form (e.g., grouping the time-to-departure into days-priors to the departure date), so the data in $\boldsymbol{T}$ can be in any time units or clustering of time periods.

Our objective is to generate a bid price matrix, $\boldsymbol{B} = [b(x,t)]$, from historical data $\boldsymbol{P}$, $\boldsymbol{T}$ and $\boldsymbol{Q}$, where $b(x,t)$ is the bid price, or the opportunity cost for remaining capacity of $x$ at the remaining time of $t$. To handle continuous quantity and capacity, $x$ is typically from a discrete set of capacity buckets aligned with operational needs, e.g. [50, 100, 200, 500, 750, 1000] representing the capacity breakpoints for a maximum weight capacity of 1000 kg.

3.2 Methodology

We extend methodology introduced in Eren et al. (2024) to account for batch arrivals with continuous quantity as described in Section 3.1. The framework is still the same composing an *observation building* step which transforms the historical booking data into a proxy of bid prices, which is then used to train a neural network to generate a model to be used to estimate future bid prices.

The modified observation building methodology is provided in Algorithm 1. The methodology is still an ex-post greedy optimization process on historical data as explained in Eren et al. (2024), ordering the prices received historically for each flight assigning them to the last $x$ remaining capacity in decreasing order. The days prior dimension is also taken care by considering only the bookings that happened for that day onward. Additionally, for air cargo case, we extend the methodology to handle continuous quantity with capacity buckets provided as $B = [b_1, b_2, \ldots, b_{|B|}]$. After sorting prices and padding with zero for any unfilled capacity, this introduces another step to calculate average unit bid price proxies corresponding to each capacity bucket.

---
**Algorithm 1** Observation Building with Continuous Quantity for Data-Driven Bid Price Generation

    **input:** historical pricing, time-to-departure and quantity data, $\boldsymbol{P}, \boldsymbol{T}$, and $\boldsymbol{Q}$
          capacity bound, $C$, set of flights $\boldsymbol{I}$, set of DCPs, $D = [d_1, d_2, \ldots, d_{|D|}]$, quantity buckets $B = [b_1, b_2, \ldots, b_{|B|}]$
      set $\tilde{P} \leftarrow [\,], \tilde{Q} \leftarrow [\,]$
      **for** $i \in \boldsymbol{I}$ **do:**
         set $\tilde{P}^i \leftarrow [\,], \tilde{Q}^i \leftarrow [\,]$
        **for** $d_j \in D$ **do:**
           set $P^i_{\leq d_j} \leftarrow \{P^i_n \mid T^i_n \leq d_j\}, \tilde{P}^i_{\leq d_j} \leftarrow [\,]$
           set $Q^i_{\leq d_j} \leftarrow \{Q^i_n \mid T^i_n \leq d_j\}, \tilde{Q}^i_{\leq d_j} \leftarrow [\,]$
           $E \leftarrow \emptyset$
           **for** $k = 1: |P^i_{\leq d_j}|$ **do:**

$$\tilde{P}^i_{\leq d_j}(k) \leftarrow \max\left(P^i_{\leq d_j} \backslash E\right)$$
# get the quantity corresponding to the maximum price
$$\tilde{Q}^i_{\leq d_j}(k) \leftarrow Q^i_{\leq d_j}\left(argmax\left(p \in P^i_{\leq d_j}: p \notin E\right)\right)$$
$$E \leftarrow E \cup \left\{\max\left(P^i_{\leq d_j} \backslash E\right)\right\}$$
**end for**
# add a zero price for the unfilled capacity
$$\tilde{P}^i_{\leq d_j}(k+1) \leftarrow 0$$
$$\tilde{Q}^i_{\leq d_j}(k+1) \leftarrow b_{|B|} - \sum_{q \in Q^i_{\leq d_j}} q$$
$$\tilde{R}^i_{\leq d_j} = \tilde{P}^i_{\leq d_j} \circ \tilde{Q}^i_{\leq d_j}$$
# calculate the cumulative revenue and quantities
**for** $k = 2: |\tilde{R}^i_{\leq d_j}|$ **do:**
$$\tilde{R}^i_{\leq d_j}(k) = \tilde{R}^i_{\leq d_j}(k-1) + \tilde{R}^i_{\leq d_j}(k)$$
$$\tilde{Q}^i_{\leq d_j}(k) = \tilde{Q}^i_{\leq d_j}(k-1) + \tilde{Q}^i_{\leq d_j}(k)$$
**end for**
# calculate unit price for each bucket by interpolation
set $\tilde{P}^i_{\leq d_j} \leftarrow [\ ]$
**for** $k = 1: |B|$ **do:**
$lower = argmax(q \in \tilde{Q}^i_{\leq d_j}: q \leq b_k)$
$upper = argmax(q \in \tilde{Q}^i_{\leq d_j}: q \geq b_k)$
$factor = (b_k - \tilde{Q}^i_{\leq d_j}(lower))/(\tilde{Q}^i_{\leq d_j}(upper) - \tilde{Q}^i_{\leq d_j}(lower))$
$interpolated = \tilde{R}^i_{\leq d_j}(lower) + factor \times (\tilde{R}^i_{\leq d_j}(upper) - \tilde{R}^i_{\leq d_j}(lower))$
set $\tilde{P}^i_{\leq d_j}(k) \leftarrow interpolated$
**end for**
$\tilde{P}^i \leftarrow [\tilde{P}^i; \tilde{P}^i_{\leq d_j}]$
**end for**
**end for**

In Algorithm 1, $|v|$ is used to denote cardinality and $v(i)$ is used to denote the $i^{th}$ element of a vector $v$. Additionally, $[A; B]$ represents row-wise concatenation of matrices $A$ and $B$, and $\circ$ is the operator for element-wise multiplication of vectors/matrices.

Once the transformed observations are built from historical booking data, the next step is feeding that input data into the estimation algorithm. Similar to the approach outlined in Eren et al. (2024) we utilize a deep neural network for estimation of bid prices, $b(x, d)$, for any given remaining capacity level, $x \in B$, and time-to-departure, $d \in D$.

Given the elimination of demand forecasting step, and the robust performance in presence of demand shocks as presented in Eren et al. (2024), the data-driven approach proves to be a good fit to address demand volatility faced in the air cargo industry. We next focus on handling of two-dimensional capacity and our findings around that, as that is another main challenge that any air cargo revenue management methodology needs to address.

### 3.2.1 Handling of Weight and Volume Dimensions in Bid Price Estimation

As air cargo capacity is two-dimensional, we would typically have two sets of bid prices as a function of weight and volume bid prices. And the historical data would compose two continuous quantities, one for weight, $W = \{W^1, W^2, \ldots, W^{|I|}\}$, and one for volume, $V = \{V^1, V^2, \ldots, V^{|I|}\}$. As per unit price would depend on the quantity used, we track the booking data in terms of total revenue $R = \{R^1, R^2, \ldots, R^{|I|}\}$, rather than the price. We present two different approaches in terms of how to handle historical *data* feeding into Algorithm 1:

1) Running Algorithm 1 with $Q = W, P = R/W, T$ for weight, and with $Q = V, P = R/V, T$ for volume, where / operator is used to represent element-wise division of the vectors in the corresponding sequences such that $p_n^i = \frac{r_n^i}{q_n^i}$.

2) Prorating historical total revenue for each booking to attribute a portion of it to weight and the remainder to volume, so that we have separate historical revenue data for weight, $R_W$, and for volume, $R_V$. Algorithm 1 then runs with $Q = W, P = R_W/W, T$ for weight, and with $Q = V, P = R_V/V, T$ for volume. We use two approaches for proration:
   a. Weight-dominated proration: Splitting each transaction's revenue in proportion to density up-to the standard air cargo density factor (1t/6m³), beyond which all the revenue is attributed to weight (Hoffmann, 2013).
   b. Volume-dominated proration: Splitting each transaction's revenue in proportion to inverse density up-to the standard air cargo inverse density factor (0.006m³/kg), beyond which all the revenue is attributed to volume.

Both approaches are followed by an independent estimation of bid prices for weight, $b_w(x, d)$, and volume, $b_v(x, d)$. We next present a more detailed explanation of proration methodologies introduced above.

3.2.1.1 Revenue proration across volume and weight

Similar to the approach followed in Hoffmann (2013), we split the revenue of each booking across weight and volume as a function of density of the shipment. For a given price, $r_n^i$, weight, $w_n^i$, and volume, $v_n^i$ for the $n^{th}$ booking of the $i^{th}$ flight, the weight-dominated proration calculates the revenue attributed to weight as

$$r_{n_w}^i = r_n^i \cdot min\left(1, \frac{w_n^i}{cw_n^i}\right), \quad (1)$$

where $cw_n^i = max\left(w_n^i, \frac{v_n^i}{0.006}\right)$ is the chargeable weight given for the booking. As a result, the revenue attributed to volume is given by

$$r_{n_v}^i = r_n^i - r_{n_w}^i. \quad (2)$$

The idea is to split the revenue in proportion to the density of the shipment for shipments with density up to the standard density factor, beyond which all the revenue gets attributed to weight. As a result, more of the revenue gets attributed to weight and weight bid prices tend to be higher, as illustrated in Figure 1.

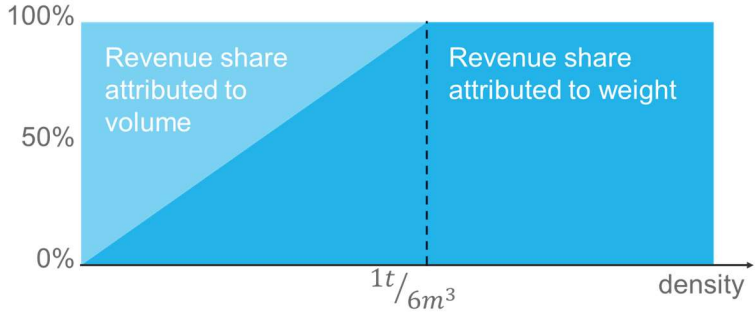

*Figure 1 Weight-dominated proration.*

For the volume-dominated proration, we reverse the methodology to attribute more revenue to volume on average, as illustrated in Figure 2. The formulae for revenue share calculation as a result change to:

$$r_{n_v}^i = r_n^i \cdot min\left(1, \frac{v_n^i/0.006}{cw_n^i}\right), \quad (3)$$

$$r_{n_w}^i = r_n^i - r_{n_v}^i. \quad (4)$$

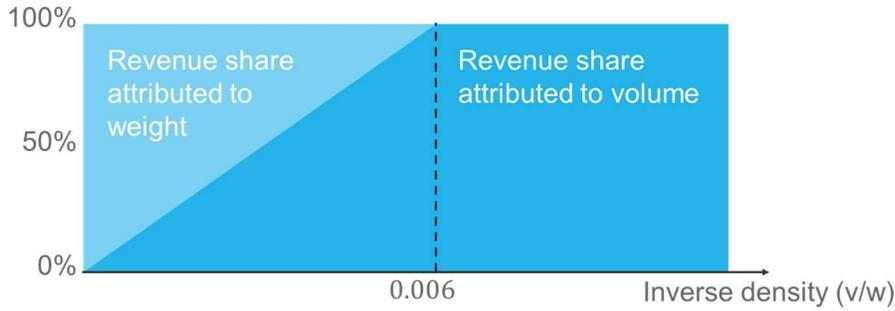

*Figure 2 Volume-dominated proration.*

3.2.2   Incorporation of Weight and Volume Bid Prices into Pricing

Next, we focus on combination of weight and volume bid prices post estimation for an incoming pricing request. Depending on the weight and volume of an incoming request, and the remaining capacity at each dimension at the time, the total bid price calculation may require unit bid prices from multiple buckets in the corresponding bid price vector. A numerical example is illustrated in Figure 3.

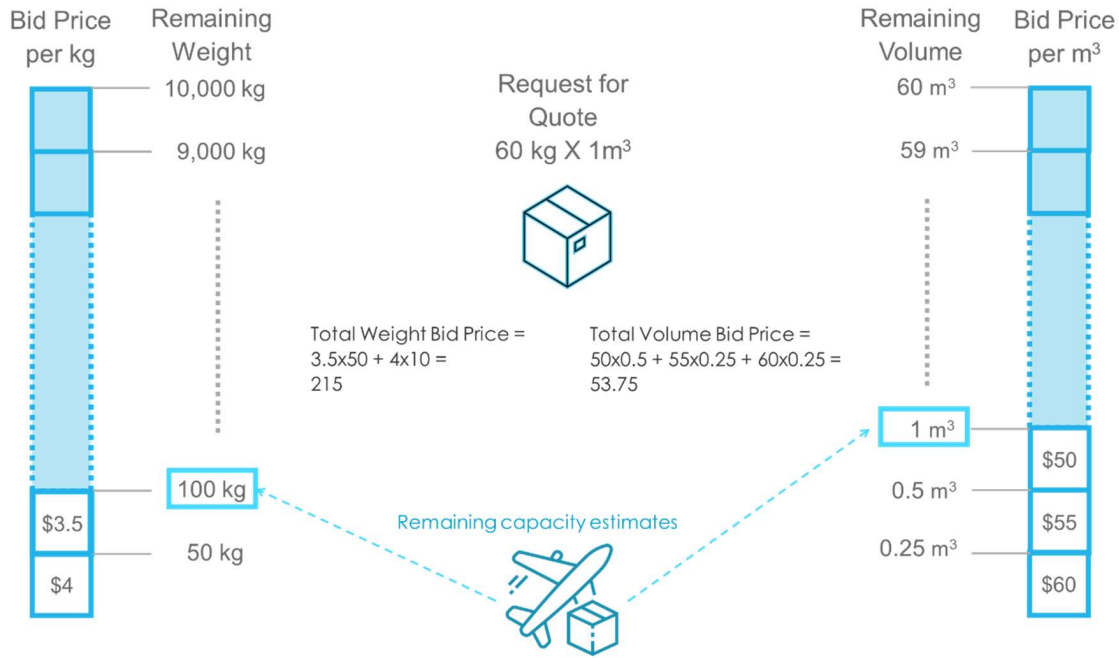

*Figure 3 Calculation and integration of weight and volume bid prices.*

Once the total weight and volume bid prices are calculated, the total bid price for the booking request can be calculated by summing the total weight and volume bid prices or getting the maximum of the two. We consider and test different scenarios for combination of weight and volume bid prices as well as handling of the historical data that feed into weight and volume estimation in Section 4.

## 4. Numerical Study and Results

In this section, we present results from our numerical study where we test the performance of our method compared to an optimal benchmark which assumes perfect knowledge of the underlying true demand model and parameters associated with it. We use simulations that mimic the single-leg air cargo setting under the context of dynamic pricing with an exponential demand model where the purchase probability at a per unit price $p$ and at time $t$ is given by

$$P_w(p,t) = e^{-\frac{(p-p_0(t))}{\alpha(t)}}, \quad (5)$$

with the mean willingness-to-pay is given by $\alpha(t)$ and $p_0(t)$ stands for the minimal price. Under this assumption and considering the batch arrivals, the optimal price is given by

$$p^*(x,t) = max\left[p_0(t), \alpha(t) + \frac{\sum_{k=1}^{x \wedge K} \pi_k \Delta^k V(x, t-\delta t)}{\sum_{k=1}^{x \wedge K} k\, \pi_k}\right], \quad (6)$$

where $\pi_k$ is the probability of a shipment with a batch size of $k = \{1, \ldots, K\}$ arriving, and $\Delta^k V(x,t) = V(x,t) - V(x-k,t)$. As defined in Eren et al. (2024), $V(x,t)$ is the value function that represents the expected revenue-to-come from future bookings given $x$ units of remaining capacity and $t$ time units to departure. The details of derivation can be found in the Appendix. We assume $p_0(t) = 0$ for the remainder of this section.

We model a Poisson arrival process where mean arrival rate $\lambda(t)$ is seasonal and both the mean arrival rate and expected WTP per unit quantity are a function of days prior as given in Figure 4. Our modeling of arrival rate is in line with the short booking horizon which is typical for air cargo. We use the assumptions from Boonekamp (2013) to model weight and inverse density with log-normal distributions (see Figure 5), aligned with the *lumpy* shipments seen in air cargo business.

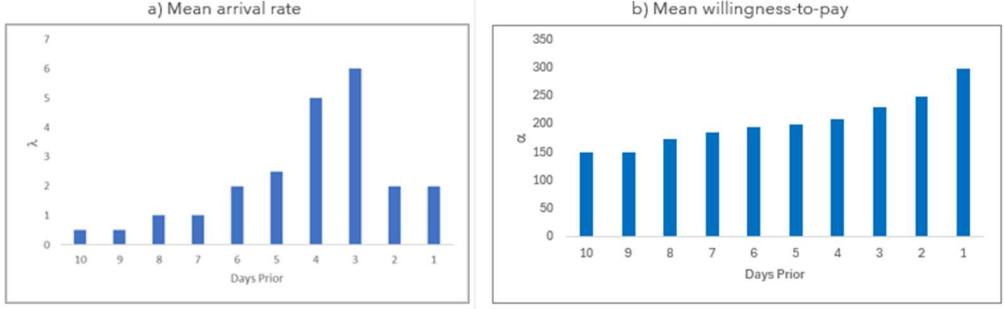

*Figure 4 a) Mean Arrival Rate vs Days Prior, b) Mean Willingness-to-Pay per Unit (=50kg) vs Days Prior.*

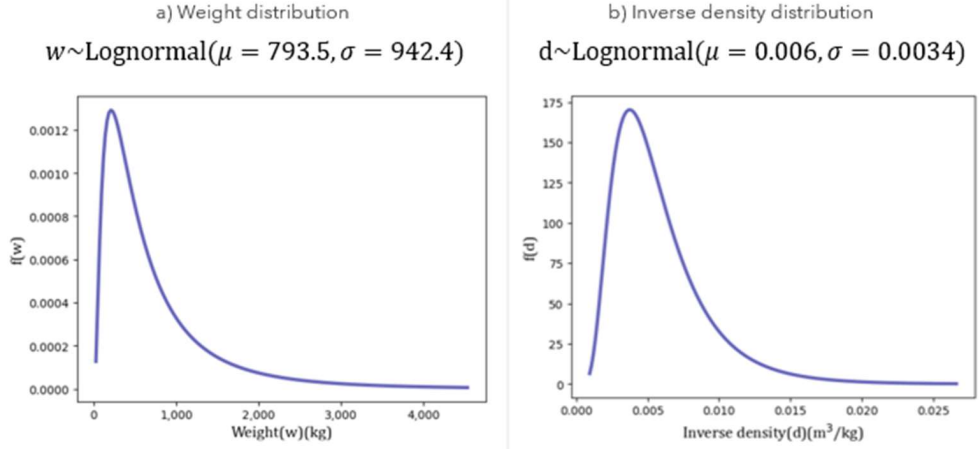

*Figure 5 a) Weight and b) Inverse Density Probability Density Functions Used in Simulations.*

We first run an initial baseline simulation with weight modeling only. We implement the optimal pricing policy which is a function of the optimal bid price and is given by

$$p^*(x,t) = \alpha(t) + \frac{\sum_{j=0}^{k-1} b^*(x-j, t-\delta t)}{k}, \qquad (7)$$

where optimal bid price $b^*(x,t)$ is defined as $b^*(x,t) = V(x,t) - V(x-1,t)$. Note that Equation (7) is a modified version of Equation (6) for an already specified batch size of an arrival. This base simulation is used to generate the historical booking data to be utilized as input to Algorithm 1 described in Section 3.2. Next, we apply Algorithm 1, feeding in pricing and time-to-departure sequences, **P, T** and **W** extracted from the simulated baseline data. Note that our methodology does not utilize any of the underlying parameters set for simulations, but only use the simulated data as an input into bid price generation, as described in Section 3.2. Using the transformed input data, $X$ and $Y$, we train the neural network model to create bid price estimate as a function of remaining capacity and DCP. We then run two sets of simulations with the same parameter setting from there on as follows, similar to the approach taken in Eren et al. (2024):

*i)* simulation with optimal bid price, which relies on the assumption of perfect knowledge of demand and WTP parameters and is an unattainable optimum in practice,

*ii)* simulation with bid prices estimated from our data-driven methodology. The pricing equation (7) still holds, replacing optimal bid price, $b^*(x,t)$, with $b^{DD}(x,t)$, thus still assuming perfect knowledge of the mean WTP, $\alpha$.

We refer the reader to Eren et al. (2024) for simulations that assess the adaptivity and robustness of our approach, where the training and test runs' underlying parameters differ. As that aspect of our methodology that makes it a good fit to address demand volatility has already been studied thoroughly, we rather focus on the testing of the weight and volume handling for the second part of this section. For this, we test four scenarios by varying proration approach and combination of weight and volume bid prices, as described in Section 3.2.1 (see Table 1 for a list of scenarios).

| Scenario | Historical price fed into algorithm | Function to combine bid prices |
|---|---|---|
| Max_original | Original | Max() |
| Sum_original | Original | Sum() |
| Sum_prorated_weight | Prorated – Weight Dominated | Sum() |
| Sum_prorated_volume | Prorated – Volume Dominated | Sum() |

*Table 1 Weight/Volume Handling Scenarios Studied in Air Cargo Simulations*

Our objective with this second set of simulations is to test how different approaches perform in terms of handling of the weight and volume bid prices. When ran independently with the same historical revenue, we test whether summing weight and volume bid prices or just taking the maximum of the two to feed into pricing results in better revenue. When algorithm is run with revenue split between weight and volume and the resulting price already, we only implement the sum of weight and volume bid prices as the total bid price, aligned with having the price input already prorated. For scenarios with proration, we test whether prorating the revenue with weight dominated vs volume dominated approach as described in Section 3.2.1.1 makes a significant difference.

We present more detailed descriptions with parameters used for both numerical studies and results next.

## 4.1 Air Cargo Baseline Simulation with Only Weight as Quantity

### 4.1.1 Simulation Setting

For baseline simulations, we model a single flight leg across 365 departure dates (365 flights, $I = \{0, \ldots, 364\}$) with a constant weight capacity set at 10,000 kg. We assume a 10-day booking horizon with a varying mean arrival rate ($\lambda_d$) and WTP ($\alpha_d$), as shown in Figure 4. We do not further group days priors and set DCPs used in Algorithm 1 as $D = [10, 9, \ldots, 1]$. We also model demand seasonality by departure date and incorporate variation by day-of-week (dow) such that the mean arrival rate for any departure date, $i$, and days prior $d$ is given by:

$$\lambda(t, d) = \lambda_d \times seasonality_{factor}(i) \times dow_{factor}[mod(i, 7)], \quad (8)$$

where $seasonality_{factor}(i) = 2.5 + \sin\left(\frac{2\pi i}{52}\right)$, and dow factors are stored in an array given by $dow_{factor} = [0.8, 1.1, 1.0, 0.95, 1.05, 1.1, 0.8]$, and $mod(x, y)$ stand for the remainder from division of $x$ by $y$.

Weight of an incoming shipment is modeled according to log-normal distribution with $\mu = 793.474$ and $\sigma = 942.37$ following the parameters used in Boonekamp (2013) as provided in Figure 5.

Once Algorithm 1 is run on the simulated data, the neural network architecture used for estimation of weight bid prices as well as the hyperparameter tuning approach are quite similar to those described in Eren et al. (2024). One major extension is the modeling of seasonality of the departure date, which is handled via Fourier series covariates. We refer the reader to Eren et al. (2024) for further details and continue to review simulation results for the sake of succinctness.

### 4.1.2. Baseline Simulation Results

Following the simulation of 365 departure dates to collect the training data for the data-driven algorithm as described in Section 4.1.1, we run two parallel simulations with the same parameter set and again for 365 departure dates to compare the revenue performance of our data-driven algorithm compared to the optimal. Figure shows the results aggregated over flight weeks. The results show an average weekly revenue gap of 1.6% from the optimal. Albeit low, especially given the unattainability of the optimal benchmark with perfect knowledge of demand parameters, the revenue gap proves to be relatively higher than what airline simulations showed in Eren et al. (2024) on average, which can be explained by the increased variation in air cargo simulations.

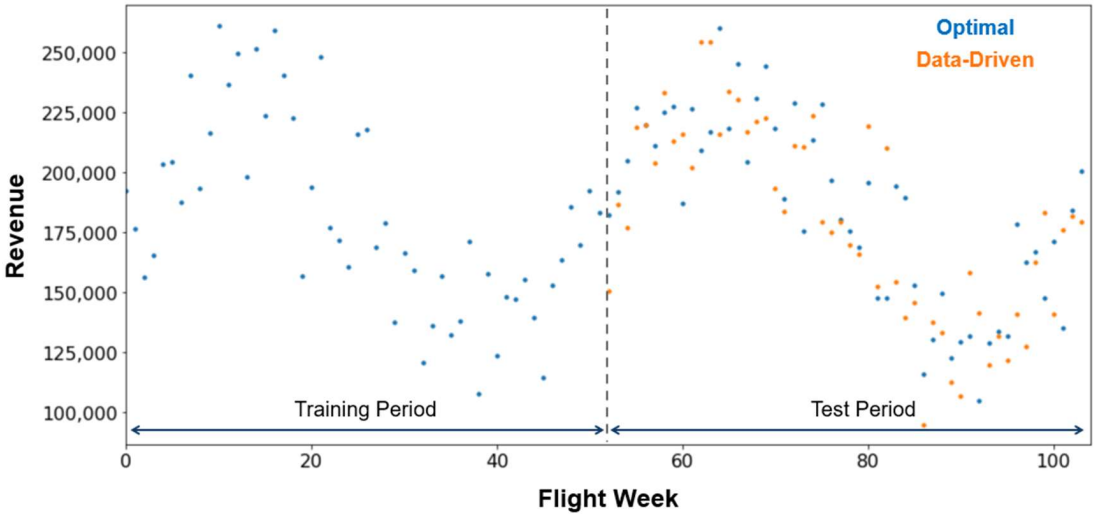

*Figure 6 Baseline Simulation Results*

4.2 Air Cargo Simulation with Both Weight and Volume

Next, we incorporate the modeling of the volume dimension into the simulation process. We model volume in simulations through inverse density, as that lets us ignore the correlation with weight and sample weight and inverse density independently. Once sampled, the volume of a shipment is simply the product of its weight and its inverse density. Inverse density of a given shipment is also assumed to be log-normally distributed (see Figure 5) with $\mu = 0.00581$ (notice the proximity to the standard inverse density factor) and $\sigma = 0.00338$. This results in a log-normally distributed volume with $\mu = 4.610$ and $\sigma = 6.879$.

We again run a single-leg simulation with 365 departure dates with similar setting as described in Section 4.1.1 extended to include volume of the shipments and a total volume capacity of 60 m$^3$. We use the training data to run Algorithm 1 with *i)* original revenue *ii)* prorated revenue (weight-dominated) and *iii)* prorated revenue (volume-dominated) to generate weight and volume bid prices with their corresponding quantities from simulation. We run a second set of parallel simulations with the corresponding bid prices from training as well as the corresponding aggregation approach to compare all the four scenarios listed in Table 1.

Figure 7 shows the total revenue aggregated over flight week per scenario. Max_original scenario outperforms all the other three scenarios by an average revenue gap of greater than 3%. In Table 2, we provide the revenue gaps for all the three underperforming scenarios with respect to the max_original scenario.

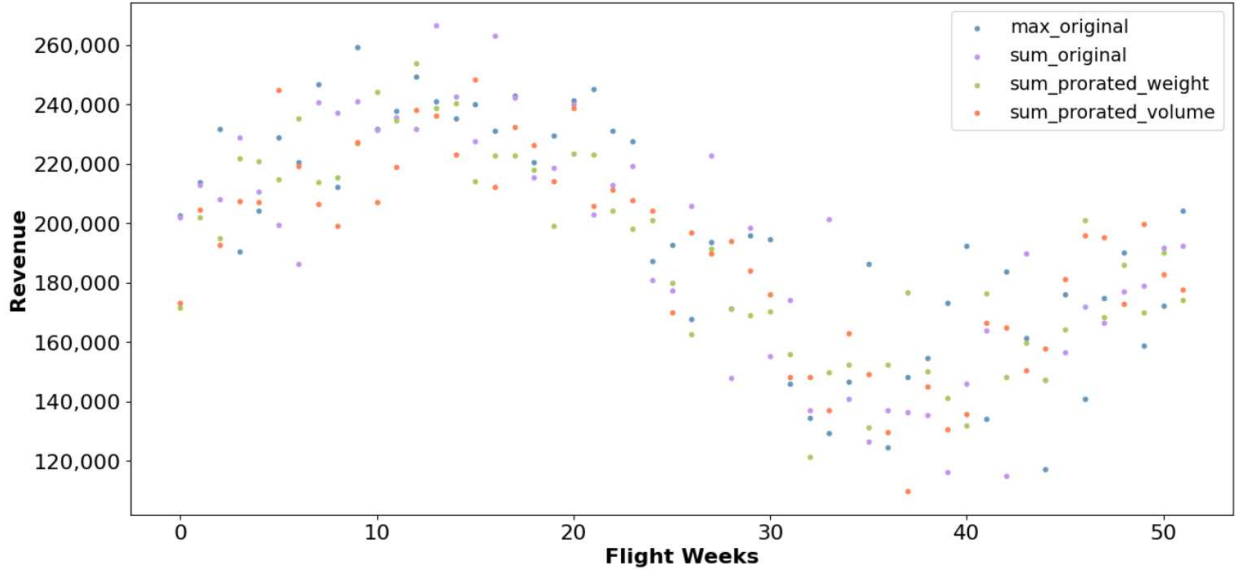

*Figure 7 Weight/Volume Simulation Results*

| Scenario | Revenue Gap |
|---|---|
| Sum_original | 3.2% |
| Sum_prorated_weight | 3.8% |
| Sum_prorated_volume | 3.7% |

*Table 2 Revenue Gaps of Scenarios Provided in Table 1 Relative to the Max_original Scenario*

The better performance of max_original scenario is in line with the fact that in practice almost always only one of the weight or volume capacity dimensions would be binding and whichever one is more constrained drives the final bid price. The results also show similar revenue performance for both weight and volume dominated proration techniques, which can be explained by the simulation setting having the total weight and volume capacity (1000 kg and 60 $m^3$) aligned with the standard air cargo inverse density factor of 0.006$m^3$/kg.

## 5. Concluding Remarks and Future Work

In this paper, we presented our data-driven bid price generation methodology (Eren et al. 2024) tailored to air cargo revenue management. Given its robustness to demand assumptions, our methodology is a great fit for air cargo where demand volatility is faced severely and continuously.

Our study of weight and volume scenarios addresses another major aspect of air cargo revenue management and provides a well-performing framework to generate and integrated weight and volume bid prices.

An extension to our work would be consideration of network problem, as both methodology and simulations so far are confined to single leg setting. Although presented in single leg context, our methodology is relevant to the network RM setting as well, with an additional step of fare proration to be added to get the historical prices to resource level prior to observation building. Extending our methodology and simulations to address network setting is a natural next step to continue proving its performance.

## Appendix: Optimal Price under Exponential Demand Assumption and Batch Arrivals

Assume that given an arrival, the batch size is $k$ with a probability $\pi_k$, $k = \{1, \ldots, K\}$. Let $x \wedge K = \min\{x, K\}$. Following the demand assumptions as presented in Section 4, and defining the state at any given time $t$ as the remaining capacity $x$ (i.e., remaining cargo weight on a flight), the dynamic programming formulation of the value function is given by

$$V(x,t) = \max_{p} \{\lambda(t)\delta t \cdot P_w(p,t) \left[\sum_{k=1}^{x \wedge K} \pi_k \, p \, k + V(x-k, t-\delta t)\right] + \left[1 - \lambda(t)\delta t \cdot P_w(p,t) + \sum_{k=x \wedge K+1}^{K} \lambda(t)\delta t \cdot P_w(p,t) \pi_k\right] V(x, t-\delta t)\}.$$

Note that the formulation simply extends to incorporate the possibility of batch arrivals and the state of remaining capacity decreasing more than 1 units in a given time slot. Rearranging terms results in Equation (6).